%% file: template.tex
\documentclass[a4paper]{article}

\usepackage{INTERSPEECH2022}
\usepackage{tikz}
\usepackage{fontawesome5}
\usepackage[hidelinks]{hyperref}
\usepackage{acronym}
\usepackage[acronym]{glossaries}
\usepackage{mathdots}
\usepackage{bm}
\usepackage{graphics}
\usepackage{amssymb}
\usepackage{float}
\usepackage{dblfloatfix}
\usepackage{multirow}
\usepackage{balance}
\usepackage{booktabs}

\input{abbreviations}
\glsdisablehyper
\usetikzlibrary{arrows.meta,patterns}
\usetikzlibrary{ipe} % ipe compatibility library

\title{VoicePrivacy 2022 System Description: Speaker Anonymization with Feature-matched F0 Trajectories}
\name{Ünal Ege Gaznepoglu$^{1,2}$, Anna Leschanowsky$^{1,2}$, Nils Peters\thanks{The International Audio Laboratories Erlangen are a joint institution of the Friedrich-Alexander-Universität Erlangen-Nürnberg and Fraunhofer IIS.}$^{1}$}
\address{
  $^1$ Friedrich-Alexander-Universität, International Audio Laboratories Erlangen, Germany\\
  $^2$ Fraunhofer IIS, Erlangen, Germany}
\email{\{uenal.ege.gaznepoglu, anna.leschanowsky\}@iis.fraunhofer.de, nils.peters@audiolabs-erlangen.de}

\begin{document}

\maketitle
\begin{abstract}
  We introduce a novel method to improve the performance of the VoicePrivacy Challenge 2022 baseline B1 variants. Among the known deficiencies of x-vector-based anonymization systems is the insufficient disentangling of the input features. In particular, the \gls{F0} trajectories, which are used for voice synthesis without any modifications. Especially in cross-gender conversion, this situation causes unnatural sounding voices, increases \glspl{WER}, and personal information leakage. Our submission overcomes this problem by synthesizing an \gls{F0} trajectory, which better harmonizes with the anonymized x-vector. We utilized a low-complexity deep neural network to estimate an appropriate \gls{F0} value per frame, using the linguistic content from the \gls{BN} and the anonymized x-vector. Our approach results in a significantly improved anonymization system and increased naturalness of the synthesized voice. Consequently, our results suggest that \gls{F0} extraction is not required for voice anonymization.
\end{abstract}

\section{Introduction}

\begin{figure*}[!bt]
    \centering
    \begingroup
\renewcommand{\baselinestretch}{1} \endlinechar=-1 \input{Figures/baseline-contributions}\endgroup \renewcommand{\baselinestretch}{1.5}
    \caption{Signal flow diagrams of the baselines B1.a (if neural vocoder is an AM-NSF), B1.b (if neural vocoder is NSF with GAN) and \texttt{joint-hifigan} (if neural vocoder is the original HiFi-GAN) together how our contribution "F0 regressor" is integrated.}
    \label{fig:baseline_contributions}
\end{figure*}
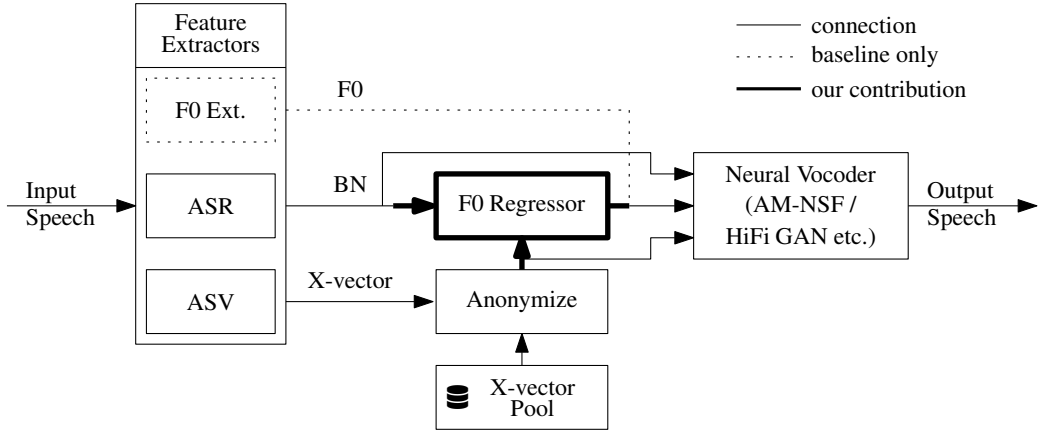

Introduction of the VoicePrivacy Challenge has stirred a multinational interest in design of voice anonymization systems. The introduced framework consists of baselines, evaluation metrics and attack models and has been utilized by researchers to improve voice anonymization. Figure~\ref{fig:baseline_contributions} depicts baseline B1 (referred to as B1.a in the current edition)~\cite{fang_speaker_2019}. Previous submissions mostly focused on changes to the individual blocks of the baselines. However, regardless of the individual modifications to this baseline by different groups, the obtained audio recordings are considered `unnatural' \cite{tomashenko_voiceprivacy_2022}. 

To improve anonymization performance as well as intelligibility, \acrshort{F0} modifications have been explored in the previous edition of the VoicePrivacy Challenge and subsequent works utilizing the challenge framework. Among the techniques investigated are creating a dictionary of F0 statistics (mean and variance) per identity and utilizing these for shifting and scaling the \acrshort{F0} trajectories \cite{champion_study_2021}, applying low-complexity DSP modifications \cite{gaznepoglu_exploring_2021} and applying functional \gls{PCA} to get speaker-dependent parts \cite{tavi_improving_2022}.
Their results show that \acrshort{F0} trajectories contribute to anonymization and modifications are promising to improve the performance of the system. 

Along the same lines, we hinted in a previous work that disentangling the features can increase the performance~\cite{gaznepoglu_exploring_2021}. In particular, \acrshort{F0}s are a complex combination of the identity of the speaker, the linguistic meaning, and the prosody, which also includes situational aspects such as emotions and speech rate \cite{johar_psychology_2016}. Many speech synthesizers, notably the \glspl{NSF}, incorporate \acrshort{F0} trajectories as a parameter to control the initial excitation, mimicking the voice cords \cite{wang_neural_2019}. Thus, data-driven parts of the architectures have relatively little control over shaping the excitation. This motivated us to investigate ways to apply a correction to the \acrshort{F0} trajectories before the synthesis such that they match the \glspl{BN} and x-vectors. Figure \ref{fig:baseline_contributions} shows how our proposal integrates into the baseline B1.

\section{Our Contributions}

\subsection{A regression DNN for F0 trajectories} %in a frame-by-frame basis

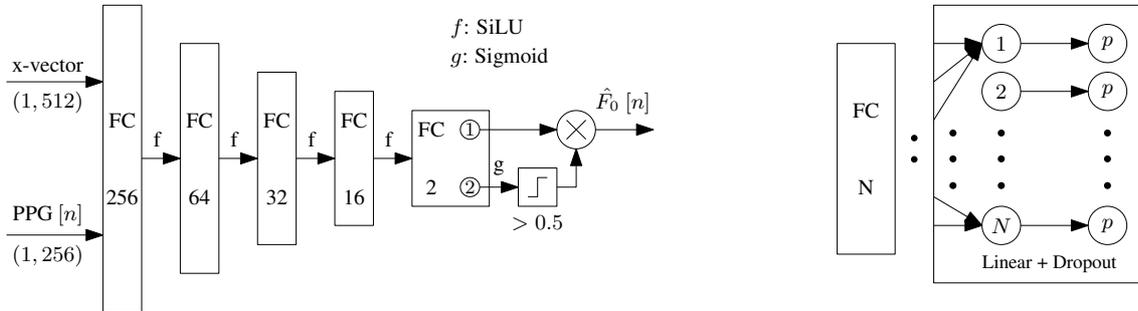
\begin{figure}[ht]
    \centering
    \scalebox{0.9}{\begingroup
\renewcommand{\baselinestretch}{1} \endlinechar=-1 \input{Figures/nn}\endgroup \renewcommand{\baselinestretch}{1.5}}
    \caption{(a) Architecture of our proposed neural network. The numbers below the expression "FC" denote the number of neurons in each layer. "FC" denotes a fully connected layer. The circles with numbers $1,2$ in the last layer denote the output of $n$th neuron in that layer (after dropout if applicable).}
    \label{fig:nn}
\end{figure}

We utilized a 3-hidden-layer \gls{DNN} (see Fig.~\ref{fig:nn}) to frame-wise predict F0 trajectories from the utterance level x-vectors and the \glspl{BN}. Internals of the so-called fully connected (FC) layer is depicted in Figure~\ref{fig:nnfc}.
F0 trajectories are predicted in logarithmic scale with a global mean-variance normalization. Two output neurons in the last layer signify the predicted pitch value $\hat{F_0}[n]$ (no activation function) and the probability of the frame signifying a voiced sound $p_v[n]$ (sigmoid activation function). According to this probability, the F0 value for the frame is either passed as is (if the probability is greater than $0.5$), or zeroed out (otherwise). The loss function $\mathcal{L}$ for a batched input is given in Equation \ref{eq:loss} where `MSE$(\cdot)$' and `BCE$(\cdot)$' denote the `mean-squared error' and `binary cross entropy with logits' as implemented by PyTorch. The variable $v$ denotes the voiced/unvoiced label of the frame and $\alpha$ denotes a trade-off parameter balancing the classification and regression tasks.

\begin{equation}
\label{eq:loss}
\mathcal{L}(F_0, \hat{F_0}) = \text{MSE}(F_0 - \hat{F_0})^2 + \alpha \text{BCE}(p_v, v),
\end{equation}

\begin{figure}[ht]
    \centering
    \scalebox{0.9}{\begingroup
\renewcommand{\baselinestretch}{1} \endlinechar=-1 \input{Figures/nnfc}\endgroup \renewcommand{\baselinestretch}{1.5}}
    \caption{ Internals of a fully connected layer. It comprises of a linear layer followed by a dropout layer, where the dropout probability is $p$. The circles with numbers $1, 2, \dots, N$ denote the number of the neuron in that layer.}
    \label{fig:nnfc}
\end{figure}

\subsubsection{Training strategies and hyperparameter optimization}
The \gls{DNN} is implemented using PyTorch \cite{paszke_pytorch_2019}, and trained using PyTorch Ignite \cite{fomin_high-level_2020}. All files in the \texttt{libri-dev-*} and \texttt{vctk-dev-*} subsets are concatenated into a single tall matrix, then a random $(90\%, 10\%)$ train-validation split is performed, allowing frames from different utterances to be present in a single batch. We use early stopping after 10 epochs without improvement and learning rate reduction (multiplication by $0.1$ after $5$ epochs without improvement in validation loss). 

OpTuna \cite{akiba_optuna_2019} tunes the learning rate \texttt{lr}, the trade-off parameter $\alpha$ and the dropout probability $p$. Optimal values obtained after 50 trials are listed in Table~\ref{tab:hyperparameters}. We found the system to perform better without dropout, thus $p = 0$.

\begin{table}[]
    \centering
    \begin{tabular}{cl}
        \toprule
        Parameter & Value \\
        \cmidrule(lr){1-1} \cmidrule(lr){2-2}
        $\alpha$ & 0.00022 \\
        \texttt{lr} & 0.0007 \\
        $p$ & 0.0\\
        \bottomrule
    \end{tabular}
    \caption{Hyperparameter values obtained using OpTuna}
    \label{tab:hyperparameters}
\end{table}

\section{Evaluation}

\subsection{Analysis of the generated F0 trajectories}

\begin{figure}[!h]
    \centering
    \input{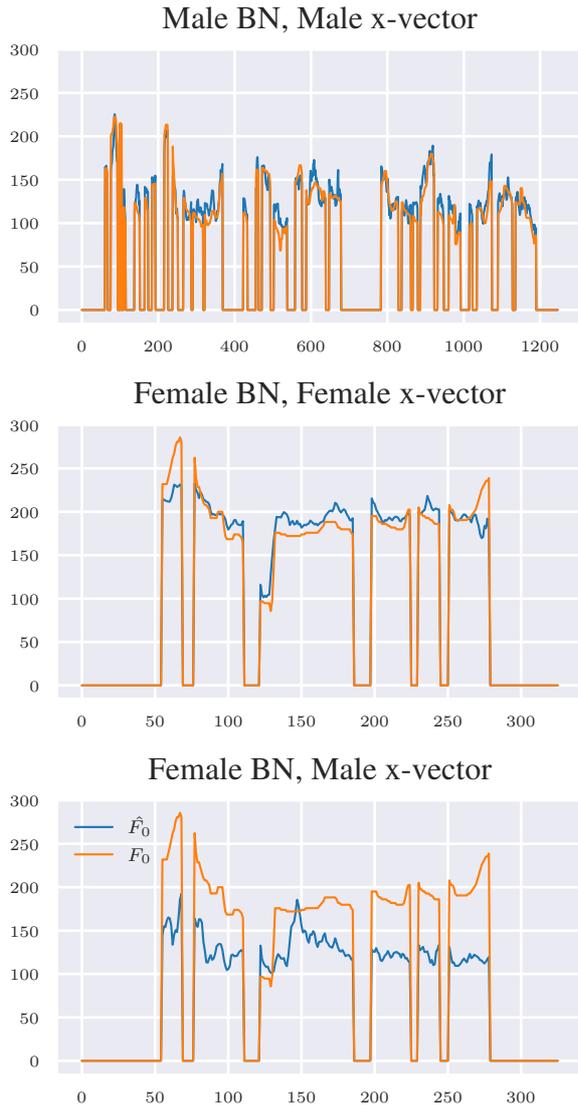}
    \caption{Ground truth F0 estimates (orange) for the input signal, obtained by YAAPT \cite{ho_accurate_2007} (F0 extractor of the B1 baselines) together with the F0 estimates obtained by our system.}
    \label{fig:reconstrutions}
\end{figure}

We verified the performance of our F0 regressor by visualizing the reconstructions for matched x-vectors and cross-gender x-vectors. The latter allows to evaluate the generalization capabilities. In Figure~\ref{fig:reconstrutions}, the F0 estimates for unaltered target and source speakers (subplots 1 and 2) as well as a cross-gender F0 conversion is given (subplot 3) for the linguistic features from the female speaker and the x-vector from the male speaker. Resulting estimated F0 trajectory has a mean shift of roughly 60~Hz and correctly identifies voiced and unvoiced frames.

While we acknowledge the necessity of thorough objective and subjective evaluation of our methodology, due to the time and space constraints we believe it is better suited as part of a different publication.

\subsection{Evaluation via challenge framework}
We executed the evaluation scripts provided by the challenge organizers. The anonymization procedure for the evaluation system training utilized the same model we obtained by training on the development subsets. As our system does not include a tunable parameter that governs the trade-off between the \gls{EER} and \gls{WER}, we submit a single set of results. As shown in Table~\ref{tab:results}, our system significantly outperforms the Baseline B1.b variant joint-hifigan in terms of \gls{EER}. The loss in \gls{WER} is negligible. Furthermore, our \gls{EER} is also significantly better than any other baseline system (c.f. \cite{tomashenko2022voiceprivacy}).
%For the VCTK conditions the \gls{WER} scores also improve. 
For every data subset the pitch correlation $\rho^{F_0}$ resides in the accepted interval $[0.3, 1]$. The voice distinctiveness $G_{VD}$ suffered some losses.

\setlength{\tabcolsep}{4pt}

\begin{table*}[!ht]
\centering
\begin{tabular}{l l c c c c c c c c c c c c}
    \toprule
    \multicolumn{1}{c}{\multirow{2}{*}{Dataset}} & \multirow{2}{*}{Sex} & \multicolumn{3}{c}{EER [\%]} & \multicolumn{3}{c}{WER [\%]}& \multicolumn{3}{c}{$\rho^{F_0}$}&\multicolumn{3}{c}{$G_{VD}$ [dB]}\\
                                 &  & B1.b & Submitted & Fixed & B1.b & Submitted & Fixed & B1.b & Submitted & Fixed & B1.b & Submitted & Fixed\\
    \cmidrule(lr){1-1} \cmidrule(lr){2-2} \cmidrule(lr){3-5} \cmidrule(lr){6-8} \cmidrule(lr){9-11} \cmidrule(lr){12-14}
    \multirow{2}{*}{{libri-dev}} & F & 16.62 & 23.86 & \textbf{24.15} & \multirow{2}{*}{\textbf{3.98}} & \multirow{2}{*}{4.12} & \multirow{2}{*}{4.13} & 0.84 & 0.82 & 0.81 & -5.86 & -6.87 & -8.95 \\
                                 & M & 5.44 & 16.15 & \textbf{16.61} & & & & 0.73 & 0.68 & 0.65 & -5.44 & -5.27 & -6.35\\
    \cmidrule(lr){1-1} \cmidrule(lr){2-2} \cmidrule(lr){3-5} \cmidrule(lr){6-8} \cmidrule(lr){9-11} \cmidrule(lr){12-14}
    \multirow{2}{*}{{vctk-dev}} & F & 7.08 & 19.71 & \textbf{16.34} & \multirow{4}{*}{\textbf{10.56}} & \multirow{4}{*}{10.36} & \multirow{4}{*}{10.62} & 0.85 & 0.86 & 0.85 & -6.57 & -6.90 & -10.70\\
                                & M & 6.55 & 17.42 & \textbf{21.69} & & & & 0.74 & 0.69 & 0.70 & -8.80 & -9.27 & -10.94\\
    %\hline
    \cmidrule(lr){1-1} \cmidrule(lr){2-2} \cmidrule(lr){3-5} \cmidrule(lr){9-11} \cmidrule(lr){12-14}
    \multirow{2}{*}{{vctk-dev-com}} & F & 8.43 & 16.57 & \textbf{11.63} & & & & 0.83 & 0.83 & 0.82 & -5.34 & -4.84 & -8.17\\
                                    & M & 9.69 & 17.38 & \textbf{22.22} & & & & 0.70 & 0.62 & 0.59 & -6.21 & -6.51 & -7.81\\
    \cmidrule(lr){1-1} \cmidrule(lr){2-2} \cmidrule(lr){3-5} \cmidrule(lr){6-8} \cmidrule(lr){9-11} \cmidrule(lr){12-14}
    \textbf{\O} (dev) & & 9.15 & 19.17 & \textbf{19.49} & \textbf{7.27} & 7.24 & 7.38 & 0.78 & 0.75 & 0.75 & -6.48 & -6.84 & -8.95\\
    \midrule
    \multirow{2}{*}{libri-test} & F & 8.39 & 22.63 & \textbf{22.99} & \multirow{2}{*}{\textbf{4.28}} & \multirow{2}{*}{4.43} & \multirow{2}{*}{4.45} &  0.83 & 0.81 & 0.82 & -5.58 & -6.16 & -7.14 \\
                                & M & 6.46 & 19.38 & \textbf{21.83} & & & & 0.68 & 0.60 & 0.59 & -5.52 & -5.54 & -5.68\\
    \cmidrule(lr){1-1} \cmidrule(lr){2-2} \cmidrule(lr){3-5} \cmidrule(lr){6-8} \cmidrule(lr){9-11} \cmidrule(lr){12-14}
    \multirow{2}{*}{vctk-test}  & F & 9.00 & 22.99 & \textbf{22.99} & \multirow{4}{*}{\textbf{10.44}} & \multirow{4}{*}{10.32} & \multirow{4}{*}{10.52} & 0.86 & 0.85 & 0.85 & -8.21 & -8.87 & -12.42\\
                                & M & 8.15 & 17.51 & \textbf{17.51} & & & & 0.75 & 0.70 & 0.70 & -8.18 & -8.81 & -10.42\\
     \cmidrule(lr){1-1} \cmidrule(lr){2-2} \cmidrule(lr){3-5} \cmidrule(lr){9-11} \cmidrule(lr){12-14}
     \multirow{2}{*}{vctk-test-com} & F & 11.56 & 19.65 & \textbf{19.65} & & & & 0.84 & 0.82 & 0.82 & -6.68 & -7.34 & -10.66 \\
                                    & M & 7.63 & 12.99 & \textbf{12.99} & & & & 0.69 & 0.63 & 0.64 & -6.11 & -6.14 & -7.43 \\
     \cmidrule(lr){1-1} \cmidrule(lr){2-2} \cmidrule(lr){3-5} \cmidrule(lr){6-8} \cmidrule(lr){9-11} \cmidrule(lr){12-14}
    \textbf{\O} (test) & & 8.10 & 20.24 & \textbf{22.07} & \textbf{7.36} & 7.38 & 7.49 & 0.78 & 0.74 & 0.74 & -6.69 & -7.14 & -8.68\\
    \bottomrule
\end{tabular}
\caption{Results from Baseline B1.b variant \texttt{joint-hifigan} taken from \cite{Results_github:2022} compared with the variant including our modifications. Better performing entries (either 'Fixed' or baseline) are highlighted for the primary metrics EER and WER. The column 'Submitted' indicates the results we have shared before the submission deadline. The column 'Fixed' indicates the results we obtained after fixing a bug within our system, counting as 'late submission'. Weighted average per challenge guidelines is denoted with \textbf{\O}.}
\label{tab:results}
\end{table*}

\section{Conclusion}
In this technical report we described our VoicePrivacy Challenge 2022 submission. Rather than extracting the F0 feature from the original speech, we proposed a novel low-complexity DNN-based F0 synthesis method which uses the linguistic content from the \glspl{BN} and the anonymized x-vector as input features. The evaluation results indicated that our method mostly preserved the WER, the pitch correlation score $\rho^{F_0}$, some reduction in the voice distinctiveness $G_{VD}$ of the baseline system, but improved the EER anonymization metric by 2.7 times. Furthermore, we observed a more natural sounding voice synthesis, especially in conditions of cross-gender voice conversion.

\balance
\bibliographystyle{IEEEtran}
\bibliography{references,localRefs}

\end{document}

%% file: abbreviations.tex
\newacronym{NSF}{NSF}{neural source-filter}
\newacronym{F0}{F0}{fundamental frequency}
\newacronym{PCA}{PCA}{principal component analysis}
\newacronym{BN}{BN}{bottleneck features}
\newacronym{TDNN}{TDNN}{time delay neural network}
\newacronym{DNN}{DNN}{deep neural network}
\newacronym{WER}{WER}{word error rate}
\newacronym{EER}{EER}{equal error rate}

%% file: Figures/baseline-contributions.tex
\tikzstyle{ipe stylesheet} = [
  ipe import,
  even odd rule,
  line join=round,
  line cap=butt,
  ipe pen normal/.style={line width=0.4},
  ipe pen heavier/.style={line width=0.8},
  ipe pen fat/.style={line width=1.2},
  ipe pen ultrafat/.style={line width=2},
  ipe pen normal,
  ipe mark normal/.style={ipe mark scale=3},
  ipe mark large/.style={ipe mark scale=5},
  ipe mark small/.style={ipe mark scale=2},
  ipe mark tiny/.style={ipe mark scale=1.1},
  ipe mark normal,
  /pgf/arrow keys/.cd,
  ipe arrow normal/.style={scale=7},
  ipe arrow large/.style={scale=10},
  ipe arrow small/.style={scale=5},
  ipe arrow tiny/.style={scale=3},
  ipe arrow normal,
  /tikz/.cd,
  ipe arrows, % update arrows
  <->/.tip = ipe normal,
  ipe dash normal/.style={dash pattern=},
  ipe dash dotted/.style={dash pattern=on 1bp off 3bp},
  ipe dash dashed/.style={dash pattern=on 4bp off 4bp},
  ipe dash dash dotted/.style={dash pattern=on 4bp off 2bp on 1bp off 2bp},
  ipe dash dash dot dotted/.style={dash pattern=on 4bp off 2bp on 1bp off 2bp on 1bp off 2bp},
  ipe dash normal,
  ipe node/.append style={font=\normalsize},
  ipe stretch normal/.style={ipe node stretch=1},
  ipe stretch normal,
  ipe opacity 10/.style={opacity=0.1},
  ipe opacity 30/.style={opacity=0.3},
  ipe opacity 50/.style={opacity=0.5},
  ipe opacity 75/.style={opacity=0.75},
  ipe opacity opaque/.style={opacity=1},
  ipe opacity opaque,
]
\definecolor{red}{rgb}{1,0,0}
\definecolor{blue}{rgb}{0,0,1}
\definecolor{green}{rgb}{0,1,0}
\definecolor{yellow}{rgb}{1,1,0}
\definecolor{orange}{rgb}{1,0.647,0}
\definecolor{gold}{rgb}{1,0.843,0}
\definecolor{purple}{rgb}{0.627,0.125,0.941}
\definecolor{gray}{rgb}{0.745,0.745,0.745}
\definecolor{brown}{rgb}{0.647,0.165,0.165}
\definecolor{navy}{rgb}{0,0,0.502}
\definecolor{pink}{rgb}{1,0.753,0.796}
\definecolor{seagreen}{rgb}{0.18,0.545,0.341}
\definecolor{turquoise}{rgb}{0.251,0.878,0.816}
\definecolor{violet}{rgb}{0.933,0.51,0.933}
\definecolor{darkblue}{rgb}{0,0,0.545}
\definecolor{darkcyan}{rgb}{0,0.545,0.545}
\definecolor{darkgray}{rgb}{0.663,0.663,0.663}
\definecolor{darkgreen}{rgb}{0,0.392,0}
\definecolor{darkmagenta}{rgb}{0.545,0,0.545}
\definecolor{darkorange}{rgb}{1,0.549,0}
\definecolor{darkred}{rgb}{0.545,0,0}
\definecolor{lightblue}{rgb}{0.678,0.847,0.902}
\definecolor{lightcyan}{rgb}{0.878,1,1}
\definecolor{lightgray}{rgb}{0.827,0.827,0.827}
\definecolor{lightgreen}{rgb}{0.565,0.933,0.565}
\definecolor{lightyellow}{rgb}{1,1,0.878}
\definecolor{black}{rgb}{0,0,0}
\definecolor{white}{rgb}{1,1,1}
\begin{tikzpicture}[ipe stylesheet]
  \draw[shift={(136, 776)}, xscale=0.875, yscale=0.8]
    (0, 0) rectangle (64, -160);
  \draw[shift={(136, 752)}, xscale=0.875]
    (0, 0)
     -- (64, 0);
  \draw[shift={(280, 640)}, yscale=0.6, ->]
    (0, 0)
     -- (0, 20);
  \draw[ipe dash dotted]
    (192.001, 736)
     -- (320, 736)
     -- (320, 700);
  \draw[shift={(280, 676)}, yscale=1.5, ipe pen ultrafat, ->]
    (0, 0)
     -- (0, 8);
  \begin{scope}[shift={(4, -8)}]
    \draw
      (136, 684) rectangle (184, 660);
    \node[ipe node, anchor=center]
       at (160, 672) {ASV};
  \end{scope}
  \begin{scope}[shift={(4, -4)}]
    \node[ipe node, anchor=center]
       at (160, 704) {ASR};
    \draw
      (136, 716) rectangle (184, 692);
  \end{scope}
  \begin{scope}[shift={(4, -4)}]
    \node[ipe node, anchor=north]
       at (160, 776) {Feature};
    \node[ipe node, anchor=north]
       at (160, 768) {Extractors};
  \end{scope}
  \begin{scope}[shift={(-8, 0)}]
    \node[ipe node, anchor=center]
       at (292, 624) {Pool};
    \node[ipe node, anchor=center]
       at (292, 632) {X-vector};
    \node[ipe node, anchor=center]
       at (264, 628) {\faDatabase};
    \draw[shift={(256, 640)}, xscale=0.8889]
      (0, 0) rectangle (72, -24);
  \end{scope}
  \begin{scope}[shift={(-8, -8)}]
    \draw[shift={(256, 684)}, xscale=1.3333]
      (0, 0) rectangle (48, -24);
    \node[ipe node, anchor=center]
       at (288, 672) {Anonymize};
  \end{scope}
  \begin{scope}[shift={(8, 0)}]
    \draw[shift={(80, 700)}, xscale=0.3, ->]
      (0, 0)
       -- (160, 0);
    \node[ipe node, anchor=south west]
       at (87.077, 701) {Input};
    \node[ipe node, anchor=north west]
       at (87.077, 699) {Speech};
  \end{scope}
  \draw[->]
    (280, 680)
     -- (328, 680)
     -- (328, 688)
     -- (344, 688);
  \begin{scope}[shift={(8, 16)}]
    \draw[shift={(336, 704.002)}, scale=1.6667]
      (0, 0) rectangle (48, -24);
    \node[ipe node, anchor=center]
       at (376, 696) {Neural Vocoder};
    \node[ipe node, anchor=center]
       at (376, 684) {(AM-NSF /};
    \node[ipe node, anchor=center]
       at (376, 672) {HiFi GAN etc.)};
  \end{scope}
  \draw[shift={(192.001, 664)}, xscale=1.7143, ->]
    (0, 0)
     -- (32, 0);
  \node[ipe node, anchor=center]
     at (216, 672) {X-vector};
  \draw[shift={(320, 700)}, xscale=0.75, ->]
    (0, 0)
     -- (32, 0);
  \draw[->]
    (228, 700)
     -- (228, 720)
     -- (328, 720)
     -- (328, 712)
     -- (344, 712);
  \draw[ipe pen ultrafat]
    (312, 700)
     -- (320, 700);
  \draw[shift={(248.001, 712)}, xscale=1.3333, ipe pen ultrafat]
    (0, 0) rectangle (48, -24);
  \node[ipe node, anchor=center]
     at (280, 700) {F0 Regressor};
  \begin{scope}
    \draw[shift={(232, 700)}, xscale=0.4643, ipe pen ultrafat, ->]
      (0, 0)
       -- (32, 0);
    \node[ipe node, anchor=center]
       at (216, 708) {BN};
    \draw
      (192, 700)
       -- (232, 700)
       -- (232, 700);
  \end{scope}
  \begin{scope}
    \draw[shift={(424, 700)}, xscale=0.3, ->]
      (0, 0)
       -- (160, 0);
    \node[ipe node, anchor=south west]
       at (431.077, 701) {Output};
    \node[ipe node, anchor=north west]
       at (431.077, 699) {Speech};
  \end{scope}
  \draw
    (360, 768)
     -- (384, 768);
  \node[ipe node, anchor=west]
     at (388, 768) {connection};
  \node[ipe node, anchor=center]
     at (216, 744) {F0};
  \draw[ipe dash dotted]
    (360, 756)
     -- (384, 756);
  \node[ipe node, anchor=west]
     at (388, 756) {baseline only};
  \begin{scope}
    \draw[ipe dash dotted]
      (140, 748) rectangle (188, 724);
    \node[ipe node, anchor=center]
       at (164, 736) {F0 Ext.};
  \end{scope}
  \draw[ipe pen fat]
    (360, 744)
     -- (384, 744);
  \node[ipe node, anchor=west]
     at (388, 744) {our contribution};
\end{tikzpicture}

%% file: Figures/nn.tex
\tikzstyle{ipe stylesheet} = [
  ipe import,
  even odd rule,
  line join=round,
  line cap=butt,
  ipe pen normal/.style={line width=0.4},
  ipe pen heavier/.style={line width=0.8},
  ipe pen fat/.style={line width=1.2},
  ipe pen ultrafat/.style={line width=2},
  ipe pen normal,
  ipe mark normal/.style={ipe mark scale=3},
  ipe mark large/.style={ipe mark scale=5},
  ipe mark small/.style={ipe mark scale=2},
  ipe mark tiny/.style={ipe mark scale=1.1},
  ipe mark normal,
  /pgf/arrow keys/.cd,
  ipe arrow normal/.style={scale=7},
  ipe arrow large/.style={scale=10},
  ipe arrow small/.style={scale=5},
  ipe arrow tiny/.style={scale=3},
  ipe arrow normal,
  /tikz/.cd,
  ipe arrows, % update arrows
  <->/.tip = ipe normal,
  ipe dash normal/.style={dash pattern=},
  ipe dash dotted/.style={dash pattern=on 1bp off 3bp},
  ipe dash dashed/.style={dash pattern=on 4bp off 4bp},
  ipe dash dash dotted/.style={dash pattern=on 4bp off 2bp on 1bp off 2bp},
  ipe dash dash dot dotted/.style={dash pattern=on 4bp off 2bp on 1bp off 2bp on 1bp off 2bp},
  ipe dash normal,
  ipe node/.append style={font=\normalsize},
  ipe stretch normal/.style={ipe node stretch=1},
  ipe stretch normal,
  ipe opacity 10/.style={opacity=0.1},
  ipe opacity 30/.style={opacity=0.3},
  ipe opacity 50/.style={opacity=0.5},
  ipe opacity 75/.style={opacity=0.75},
  ipe opacity opaque/.style={opacity=1},
  ipe opacity opaque,
]
\definecolor{red}{rgb}{1,0,0}
\definecolor{blue}{rgb}{0,0,1}
\definecolor{green}{rgb}{0,1,0}
\definecolor{yellow}{rgb}{1,1,0}
\definecolor{orange}{rgb}{1,0.647,0}
\definecolor{gold}{rgb}{1,0.843,0}
\definecolor{purple}{rgb}{0.627,0.125,0.941}
\definecolor{gray}{rgb}{0.745,0.745,0.745}
\definecolor{brown}{rgb}{0.647,0.165,0.165}
\definecolor{navy}{rgb}{0,0,0.502}
\definecolor{pink}{rgb}{1,0.753,0.796}
\definecolor{seagreen}{rgb}{0.18,0.545,0.341}
\definecolor{turquoise}{rgb}{0.251,0.878,0.816}
\definecolor{violet}{rgb}{0.933,0.51,0.933}
\definecolor{darkblue}{rgb}{0,0,0.545}
\definecolor{darkcyan}{rgb}{0,0.545,0.545}
\definecolor{darkgray}{rgb}{0.663,0.663,0.663}
\definecolor{darkgreen}{rgb}{0,0.392,0}
\definecolor{darkmagenta}{rgb}{0.545,0,0.545}
\definecolor{darkorange}{rgb}{1,0.549,0}
\definecolor{darkred}{rgb}{0.545,0,0}
\definecolor{lightblue}{rgb}{0.678,0.847,0.902}
\definecolor{lightcyan}{rgb}{0.878,1,1}
\definecolor{lightgray}{rgb}{0.827,0.827,0.827}
\definecolor{lightgreen}{rgb}{0.565,0.933,0.565}
\definecolor{lightyellow}{rgb}{1,1,0.878}
\definecolor{black}{rgb}{0,0,0}
\definecolor{white}{rgb}{1,1,1}
\begin{tikzpicture}[ipe stylesheet]
  \draw[shift={(44, 800)}, xscale=0.625, ->]
    (0, 0)
     -- (64, 0);
  \draw[shift={(44, 736)}, xscale=0.625, ->]
    (0, 0)
     -- (64, 0);
  \node[ipe node, anchor=south east, font=\small]
     at (76, 804) {x-vector};
  \node[ipe node, anchor=south east, font=\small]
     at (76, 740) {PPG $[n]$};
  \node[ipe node, anchor=north east, font=\small]
     at (76, 796) {$(1, 512)$};
  \node[ipe node, anchor=north east, font=\small]
     at (76, 732) {$(1, 256)$};
  \draw[shift={(240, 780)}, xscale=0.5, ->]
    (0, 0)
     -- (64, 0);
  \node[ipe node, anchor=west, font=\small]
     at (288, 792) {$\hat{F_0}$ $[n]$};
  \draw[->]
    (272, 756)
     -- (280, 756)
     -- (280, 772);
  \draw[shift={(288, 780)}, xscale=1.5, ->]
    (0, 0)
     -- (16, 0);
  \begin{scope}[shift={(-120, 44)}]
    \draw
      (400, 736) circle[radius=8];
    \draw
      (396, 740)
       -- (404, 732);
    \draw
      (404, 740)
       -- (396, 732);
  \end{scope}
  \begin{scope}[shift={(16.7993, 0)}, xscale=0.8]
    \draw[shift={(103.999, 768)}, xscale=0.8333, ->]
      (0, 0)
       -- (24, 0);
    \node[ipe node, anchor=center]
       at (112, 776) {f};
  \end{scope}
  \begin{scope}[shift={(48.7993, 0)}, xscale=0.8]
    \draw[shift={(103.999, 768)}, xscale=0.8333, ->]
      (0, 0)
       -- (24, 0);
    \node[ipe node, anchor=center]
       at (112, 776) {f};
  \end{scope}
  \begin{scope}[shift={(80.799, 0)}, xscale=0.8]
    \draw[shift={(103.999, 768)}, xscale=0.8333, ->]
      (0, 0)
       -- (24, 0);
    \node[ipe node, anchor=center]
       at (112, 776) {f};
  \end{scope}
  \draw[shift={(212, 788)}, xscale=0.6667, yscale=0.25]
    (0, 0) rectangle (48, -160);
  \node[ipe node, anchor=center, font=\small]
     at (220, 780) {FC};
  \node[ipe node, anchor=center, font=\small]
     at (220, 756) {2};
  \draw[shift={(240, 756)}, xscale=0.25, ->]
    (0, 0)
     -- (64, 0);
  \node[ipe node, anchor=center]
     at (248, 764) {g};
  \begin{scope}[shift={(-28, 0)}]
    \draw[shift={(224, 768)}, xscale=0.6667, ->]
      (0, 0)
       -- (24, 0);
    \node[ipe node, anchor=center]
       at (230.4, 776) {f};
  \end{scope}
  \node[ipe node]
     at (228, 820) {$f$: {SiLU}};
  \node[ipe node]
     at (228, 808) {$g$: Sigmoid};
  \node[ipe node, anchor=north, font=\small]
     at (264, 744) {$> 0.5$};
  \begin{scope}[shift={(-28, -4)}]
    \draw
      (264, 760) circle[radius=4];
    \node[ipe node, anchor=center, font=\footnotesize]
       at (264, 760) {$2$};
  \end{scope}
  \begin{scope}[shift={(-28, 4)}]
    \draw
      (264, 776) circle[radius=4];
    \node[ipe node, anchor=center, font=\footnotesize]
       at (264, 776) {$1$};
  \end{scope}
  \begin{scope}[shift={(-32, -4)}]
    \draw[shift={(288, 768.001)}, yscale=0.6667]
      (0, 0) rectangle (16, -24);
    \draw[shift={(292, 756)}, yscale=0.5]
      (0, 0)
       -- (4, 0)
       -- (4, 16)
       -- (8, 16);
  \end{scope}
  \draw[shift={(148, 804)}, xscale=0.3333, yscale=0.45]
    (0, 0) rectangle (48, -160);
  \node[ipe node, anchor=center, font=\small]
     at (156, 784) {FC};
  \node[ipe node, anchor=center, font=\small]
     at (156, 752) {32};
  \begin{scope}[shift={(-8, 0)}]
    \draw[shift={(124, 816)}, xscale=0.3333, yscale=0.6]
      (0, 0) rectangle (48, -160);
    \node[ipe node, anchor=center, font=\small]
       at (132, 784) {FC};
    \node[ipe node, anchor=center, font=\small]
       at (132, 752) {64};
  \end{scope}
  \begin{scope}
    \draw[shift={(84, 832)}, xscale=0.3333, yscale=0.8]
      (0, 0) rectangle (48, -160);
    \node[ipe node, anchor=center, font=\small]
       at (92, 784) {FC};
    \node[ipe node, anchor=center, font=\small]
       at (92, 752) {256};
  \end{scope}
  \begin{scope}[shift={(-24, 0)}]
    \draw[shift={(204, 796)}, xscale=0.3333, yscale=0.35]
      (0, 0) rectangle (48, -160);
    \node[ipe node, anchor=center, font=\small]
       at (212, 784) {FC};
    \node[ipe node, anchor=center, font=\small]
       at (212, 752) {16};
  \end{scope}
\end{tikzpicture}

%% file: Figures/nnfc.tex
\tikzstyle{ipe stylesheet} = [
  ipe import,
  even odd rule,
  line join=round,
  line cap=butt,
  ipe pen normal/.style={line width=0.4},
  ipe pen heavier/.style={line width=0.8},
  ipe pen fat/.style={line width=1.2},
  ipe pen ultrafat/.style={line width=2},
  ipe pen normal,
  ipe mark normal/.style={ipe mark scale=3},
  ipe mark large/.style={ipe mark scale=5},
  ipe mark small/.style={ipe mark scale=2},
  ipe mark tiny/.style={ipe mark scale=1.1},
  ipe mark normal,
  /pgf/arrow keys/.cd,
  ipe arrow normal/.style={scale=7},
  ipe arrow large/.style={scale=10},
  ipe arrow small/.style={scale=5},
  ipe arrow tiny/.style={scale=3},
  ipe arrow normal,
  /tikz/.cd,
  ipe arrows, % update arrows
  <->/.tip = ipe normal,
  ipe dash normal/.style={dash pattern=},
  ipe dash dotted/.style={dash pattern=on 1bp off 3bp},
  ipe dash dashed/.style={dash pattern=on 4bp off 4bp},
  ipe dash dash dotted/.style={dash pattern=on 4bp off 2bp on 1bp off 2bp},
  ipe dash dash dot dotted/.style={dash pattern=on 4bp off 2bp on 1bp off 2bp on 1bp off 2bp},
  ipe dash normal,
  ipe node/.append style={font=\normalsize},
  ipe stretch normal/.style={ipe node stretch=1},
  ipe stretch normal,
  ipe opacity 10/.style={opacity=0.1},
  ipe opacity 30/.style={opacity=0.3},
  ipe opacity 50/.style={opacity=0.5},
  ipe opacity 75/.style={opacity=0.75},
  ipe opacity opaque/.style={opacity=1},
  ipe opacity opaque,
]
\definecolor{red}{rgb}{1,0,0}
\definecolor{blue}{rgb}{0,0,1}
\definecolor{green}{rgb}{0,1,0}
\definecolor{yellow}{rgb}{1,1,0}
\definecolor{orange}{rgb}{1,0.647,0}
\definecolor{gold}{rgb}{1,0.843,0}
\definecolor{purple}{rgb}{0.627,0.125,0.941}
\definecolor{gray}{rgb}{0.745,0.745,0.745}
\definecolor{brown}{rgb}{0.647,0.165,0.165}
\definecolor{navy}{rgb}{0,0,0.502}
\definecolor{pink}{rgb}{1,0.753,0.796}
\definecolor{seagreen}{rgb}{0.18,0.545,0.341}
\definecolor{turquoise}{rgb}{0.251,0.878,0.816}
\definecolor{violet}{rgb}{0.933,0.51,0.933}
\definecolor{darkblue}{rgb}{0,0,0.545}
\definecolor{darkcyan}{rgb}{0,0.545,0.545}
\definecolor{darkgray}{rgb}{0.663,0.663,0.663}
\definecolor{darkgreen}{rgb}{0,0.392,0}
\definecolor{darkmagenta}{rgb}{0.545,0,0.545}
\definecolor{darkorange}{rgb}{1,0.549,0}
\definecolor{darkred}{rgb}{0.545,0,0}
\definecolor{lightblue}{rgb}{0.678,0.847,0.902}
\definecolor{lightcyan}{rgb}{0.878,1,1}
\definecolor{lightgray}{rgb}{0.827,0.827,0.827}
\definecolor{lightgreen}{rgb}{0.565,0.933,0.565}
\definecolor{lightyellow}{rgb}{1,1,0.878}
\definecolor{black}{rgb}{0,0,0}
\definecolor{white}{rgb}{1,1,1}
\begin{tikzpicture}[ipe stylesheet]
  \draw[shift={(151.999, 784)}, xscale=1.8333, yscale=0.725]
    (0, 0) rectangle (48, -160);
  \node[ipe node, anchor=center, font=\footnotesize]
     at (200, 676) {Linear + Dropout};
  \node[ipe node, anchor=center]
     at (224, 728) {\scalebox{3}{$\vdots$}};
  \node[ipe node, anchor=center]
     at (224, 768) {$p$};
  \node[ipe node, anchor=center]
     at (224, 748) {$p$};
  \node[ipe node, anchor=center]
     at (224, 692) {$p$};
  \node[ipe node, anchor=center]
     at (180, 768) {$1$};
  \node[ipe node, anchor=center]
     at (180, 748) {$2$};
  \node[ipe node, anchor=center]
     at (180, 692) {$N$};
  \node[ipe node, anchor=center]
     at (180, 728) {\scalebox{3}{$\vdots$}};
  \draw
    (180, 768) circle[radius=8];
  \draw
    (224, 768) circle[radius=8];
  \draw
    (180, 748) circle[radius=8];
  \draw
    (224, 748) circle[radius=8];
  \draw
    (180, 692) circle[radius=8];
  \draw
    (224, 692) circle[radius=8];
  \draw[->]
    (188, 768)
     -- (216, 768);
  \draw[->]
    (188, 748)
     -- (216, 748);
  \draw[->]
    (188, 692)
     -- (216, 692);
  \draw[->]
    (152, 768)
     -- (172, 768);
  \draw[shift={(152, 752)}, yscale=0.8, ->]
    (0, 0)
     -- (20, 20);
  \draw[->]
    (152, 736)
     -- (172, 768);
  \node[ipe node, anchor=center]
     at (160, 728) {\scalebox{3}{$\vdots$}};
  \draw[->]
    (152, 692)
     -- (172, 692);
  \draw[->]
    (152, 704)
     -- (172, 692);
  \node[ipe node, anchor=center, font=\Huge]
     at (144, 724) {:};
  \begin{scope}[shift={(-256, -48)}]
    \draw[shift={(368, 816)}, xscale=0.5, yscale=0.55]
      (0, 0) rectangle (48, -160);
    \node[ipe node, anchor=center]
       at (380, 788) {FC};
    \node[ipe node, anchor=center]
       at (380, 756) {N};
  \end{scope}
\end{tikzpicture}

%% file: template.bbl
\begin{thebibliography}{10}
\providecommand{\url}[1]{#1}
\csname url@samestyle\endcsname
\providecommand{\newblock}{\relax}
\providecommand{\bibinfo}[2]{#2}
\providecommand{\BIBentrySTDinterwordspacing}{\spaceskip=0pt\relax}
\providecommand{\BIBentryALTinterwordstretchfactor}{4}
\providecommand{\BIBentryALTinterwordspacing}{\spaceskip=\fontdimen2\font plus
\BIBentryALTinterwordstretchfactor\fontdimen3\font minus
  \fontdimen4\font\relax}
\providecommand{\BIBforeignlanguage}[2]{{%
\expandafter\ifx\csname l@#1\endcsname\relax
\typeout{** WARNING: IEEEtran.bst: No hyphenation pattern has been}%
\typeout{** loaded for the language `#1'. Using the pattern for}%
\typeout{** the default language instead.}%
\else
\language=\csname l@#1\endcsname
\fi
#2}}
\providecommand{\BIBdecl}{\relax}
\BIBdecl

\bibitem{fang_speaker_2019}
\BIBentryALTinterwordspacing
F.~Fang, X.~Wang, J.~Yamagishi, I.~Echizen, M.~Todisco, N.~Evans, and J.-F.
  Bonastre, ``Speaker {Anonymization} {Using} {X}-vector and {Neural}
  {Waveform} {Models},'' \emph{arXiv:1905.13561 [cs, eess, stat]}, May 2019,
  00061 arXiv: 1905.13561. [Online]. Available:
  \url{http://arxiv.org/abs/1905.13561}
\BIBentrySTDinterwordspacing

\bibitem{tomashenko_voiceprivacy_2022}
\BIBentryALTinterwordspacing
N.~Tomashenko, X.~Wang, E.~Vincent, J.~Patino, B.~M.~L. Srivastava, P.-G. Noé,
  A.~Nautsch, N.~Evans, J.~Yamagishi, B.~O'Brien, A.~Chanclu, J.-F. Bonastre,
  M.~Todisco, and M.~Maouche, ``\BIBforeignlanguage{en}{The {VoicePrivacy} 2020
  {Challenge}: {Results} and findings},''
  \emph{\BIBforeignlanguage{en}{Computer Speech \& Language}}, vol.~74, p.
  101362, Jul. 2022, 00015 arXiv:2109.00648 [cs, eess]. [Online]. Available:
  \url{http://arxiv.org/abs/2109.00648}
\BIBentrySTDinterwordspacing

\bibitem{champion_study_2021}
\BIBentryALTinterwordspacing
P.~Champion, D.~Jouvet, and A.~Larcher, ``\BIBforeignlanguage{en}{A {Study} of
  {F0} {Modification} for {X}-{Vector} {Based} {Speech} {Pseudonymization}
  {Across} {Gender}},'' \emph{\BIBforeignlanguage{en}{arXiv:2101.08478 [cs,
  eess]}}, Jan. 2021, 00009 arXiv: 2101.08478. [Online]. Available:
  \url{http://arxiv.org/abs/2101.08478}
\BIBentrySTDinterwordspacing

\bibitem{gaznepoglu_exploring_2021}
\BIBentryALTinterwordspacing
U.~E. Gaznepoglu and N.~Peters, ``\BIBforeignlanguage{en}{Exploring the
  {Importance} of {F0} {Trajectories} for {Speaker} {Anonymization} using
  {X}-vectors and {Neural} {Waveform} {Models}},''
  \emph{\BIBforeignlanguage{en}{Workshop on Machine Learning in Speech and
  Language Processing 2021}}, Sep. 2021, 00002. [Online]. Available:
  \url{https://arxiv.org/abs/2110.06887v1}
\BIBentrySTDinterwordspacing

\bibitem{tavi_improving_2022}
\BIBentryALTinterwordspacing
L.~Tavi, T.~Kinnunen, and R.~G. Hautamäki, ``\BIBforeignlanguage{en}{Improving
  speaker de-identification with functional data analysis of f0
  trajectories},'' \emph{\BIBforeignlanguage{en}{Speech Communication}}, vol.
  140, pp. 1--10, May 2022, 00000 arXiv:2203.16738 [cs, eess]. [Online].
  Available: \url{http://arxiv.org/abs/2203.16738}
\BIBentrySTDinterwordspacing

\bibitem{johar_psychology_2016}
\BIBentryALTinterwordspacing
S.~Johar, ``\BIBforeignlanguage{en}{Psychology of {Voice}},'' in
  \emph{\BIBforeignlanguage{en}{Emotion, {Affect} and {Personality} in
  {Speech}: {The} {Bias} of {Language} and {Paralanguage}}}, ser.
  {SpringerBriefs} in {Electrical} and {Computer} {Engineering}, S.~Johar,
  Ed.\hskip 1em plus 0.5em minus 0.4em\relax Cham: Springer International
  Publishing, 2016, pp. 9--15, 00014. [Online]. Available:
  \url{https://doi.org/10.1007/978-3-319-28047-9_2}
\BIBentrySTDinterwordspacing

\bibitem{wang_neural_2019}
\BIBentryALTinterwordspacing
X.~Wang and J.~Yamagishi, ``Neural {Harmonic}-plus-{Noise} {Waveform} {Model}
  with {Trainable} {Maximum} {Voice} {Frequency} for {Text}-to-{Speech}
  {Synthesis},'' \emph{arXiv:1908.10256 [cs, eess]}, Aug. 2019, 00027 arXiv:
  1908.10256. [Online]. Available: \url{http://arxiv.org/abs/1908.10256}
\BIBentrySTDinterwordspacing

\bibitem{paszke_pytorch_2019}
A.~Paszke, S.~Gross, F.~Massa, A.~Lerer, J.~Bradbury, G.~Chanan, T.~Killeen,
  Z.~Lin, N.~Gimelshein, L.~Antiga, A.~Desmaison, A.~Kopf, E.~Yang, Z.~DeVito,
  M.~Raison, A.~Tejani, S.~Chilamkurthy, B.~Steiner, L.~Fang, J.~Bai, and
  S.~Chintala, ``{PyTorch}: {An} {Imperative} {Style}, {High}-{Performance}
  {Deep} {Learning} {Library},'' in \emph{Advances in neural information
  processing systems ({NeurIPS})}, Vancouver, Canada, 2019, p.~12, 17858.

\bibitem{fomin_high-level_2020}
\BIBentryALTinterwordspacing
V.~Fomin, J.~Anmol, S.~Desroziers, J.~Kriss, and A.~Tejani, ``High-level
  library to help with training neural networks in {PyTorch},'' 2020, 00014
  Publication Title: GitHub repository. [Online]. Available:
  \url{https://github.com/pytorch/ignite}
\BIBentrySTDinterwordspacing

\bibitem{akiba_optuna_2019}
\BIBentryALTinterwordspacing
T.~Akiba, S.~Sano, T.~Yanase, T.~Ohta, and M.~Koyama, ``Optuna: {A}
  {Next}-generation {Hyperparameter} {Optimization} {Framework},'' Jul. 2019,
  01169 Number: arXiv:1907.10902 arXiv:1907.10902 [cs, stat]. [Online].
  Available: \url{http://arxiv.org/abs/1907.10902}
\BIBentrySTDinterwordspacing

\bibitem{ho_accurate_2007}
K.~C. Ho and M.~Sun, ``An {Accurate} {Algebraic} {Closed}-{Form} {Solution} for
  {Energy}-{Based} {Source} {Localization},'' \emph{IEEE Transactions on Audio,
  Speech, and Language Processing}, vol.~15, no.~8, pp. 2542--2550, Nov. 2007,
  00111 Conference Name: IEEE Transactions on Audio, Speech, and Language
  Processing.

\bibitem{tomashenko2022voiceprivacy}
N.~Tomashenko, X.~Wang, X.~Miao, H.~Nourtel, P.~Champion, M.~Todisco,
  E.~Vincent, N.~Evans, J.~Yamagishi, and J.~F. Bonastre, ``The voiceprivacy
  2022 challenge evaluation plan,'' \emph{arXiv preprint arXiv:2203.12468},
  2022.

\bibitem{Results_github:2022}
\BIBentryALTinterwordspacing
Baseline results for \texttt{joint-hifigan}. (last accessed 2022-07-31).
  [Online]. Available:
  \url{https://github.com/Voice-Privacy-Challenge/Voice-Privacy-Challenge-2022/blob/master/baseline/results/RESULTS_summary_tts_joint_hifigan}
\BIBentrySTDinterwordspacing

\end{thebibliography}
